\documentclass[aps,superscriptaddress,showpacs]{revtex4}
\usepackage{amsmath}
\usepackage{graphicx}
\begin{document}
\title{Relativistic Laser-Plasma Interactions in the Quantum Regime}
\author{Bengt Eliasson}
\affiliation{Department of Physics, Ume{\aa} University,
SE-901 87 Ume{\aa}, Sweden}
\affiliation{Institut f\"ur Theoretische Physik,
Fakult\"at f\"ur Physik und Astronomie,
Ruhr--Universit\"at Bochum, D-44780 Bochum, Germany}
\author{P. K. Shukla}
\affiliation{RUB International Chair, International Centre for Advanced Studies in Physical Sciences,
Fakult\"at f\"ur Physik und Astronomie, Ruhr--Universit\"at Bochum, D-44780 Bochum, Germany}
\received{14 October 2010}
\begin{abstract}
We investigate the nonlinear interaction between a relativistically strong laser beam and
a plasma in the quantum regime. The collective behavior of the electrons is modeled by
a Klein-Gordon equation, which is nonlinearly coupled with the electromagnetic wave
through the Maxwell and Poisson equations. This allows us to study the nonlinear
interaction between arbitrarily large amplitude electromagnetic waves and a quantum plasma.
We have used our system of nonlinear equations to study theoretically the parametric instabilities
involving stimulated Raman scattering and modulational instabilities. A model for quasi-steady
state propagating electromagnetic wavepackets is also derived, and which shows the possibility
of localized solitary structures in the quantum plasma. Numerical simulations demonstrate
the collapse and acceleration of the electrons in the nonlinear stage of the modulational
instability, as well as the possibility of wake-field acceleration of the electrons to
relativistic speeds by short laser pulses at nanometer length scales. The study has importance
for the nonlinear interaction between a super-intense X-ray laser light and a solid-density plasma,
where quantum effects are important.
\end{abstract}

\maketitle

\section{Introduction}

With the advent of the X-ray free-electron lasers \cite{Hand09} there
are new possibilities to explore matter on atomic and single molecule
levels. On these length scales, of the order of a few {\AA}ngstr\"om, quantum
effects play an important role in the dynamics of the electrons.
Quantum effects have been measured experimentally both in the degenerate electron gas in
metals and in warm dense matters \cite{Glenzer}.  It has also been found that quantum
mechanical effects must be taken into account in intense laser-solid density plasma
interaction experiments \cite{Andreev,Bulanov,MarklundShukla}. The interaction of large
amplitude electromagnetic waves with the plasma can lead to various parametric
instabilities \cite{Drake74,Sharma83,Murtaza84}. At laser intensities around $10^{19}\,\mathrm{W/cm}^2$
and above, the nonlinearity associated with relativistic electron mass increase in short laser pulses
plays a significant role. Furthermore, the relativistic ponderomotive
force \cite{Shukla} of intense laser pulses produces density modifications.
Thus, in a classical plasma, nonlinear effects associated with relativistic electron mass
increase and relativistic ponderomotive  force very important, since
they provide the possibility of the modulational instability \cite{McKinstrie89,Tsintsadze91}
followed by a compression and localization of intense electromagnetic waves.
In addition to the modulational instability, there are relativistic Raman forward and
backward scattering instabilities \cite{McKinstrie92,Sakharov94,Guerin95,Adam00}
and the two-plasmon decay \cite{Quesnel97} instability that lead to strong collisionless
heating of the plasma in the relativistic regime. The parametric instabilities of
intense electromagnetic waves in magnetized plasmas have also been
investigated \cite{Stenflo76,Stenflo80,Stenflo81}.

However, for intense electromagnetic waves interacting with the plasma in the
X-ray and $\gamma$-ray regimes, both relativistic and quantum effects must me
taken into account on equal footing. Accordingly, in this paper, we present a
simple nonlinear model, based on the Klein-Gordon (KG) equation coupled with the
Maxwell equations that are capable of treating both the relativistic and quantum effects.
Our work has applications in laboratories \cite{Glenzer,Malkin07}, in quantum free electron
laser systems \cite{Serbeto08,Serbeto09,Piovella08}, as well as in astrophysical
settings \cite{Chabrier} where white dwarf cores \cite{Coe} and neutron stars \cite{Hurley}
are strong sources of x-rays and $\gamma$-rays.

The manuscript is organized as follows. In Sec. II, we present our mathematical model for
the coupled KG and Maxwell equations, exhibiting nonlinear interactions between
relativistic electrons and electromagnetic fields. Linear properties of the electrostatic
and electromagnetic waves are discussed in Sec. III. Section IV shows hoe our governing
equations lead to the wave equation that reveals the phenomena of relativistic self-focusing
and relativistic self-induced transparency of electromagnetic waves. Section V is concerned
with the theoretical and numerical investigations of the relativistic parametric instabilities
in the quantum regime. Section VI deals with relativistic optical solitary waves. The nonlinear
dynamics of interacting intense localized electromagnetic pulses, as well as the new phenomena
of the formation of nonlinear Bernstein-Greene-Krushkal (BGK)-like modes and associated electron
acceleration are described in Sec. VII. Section VIII contains a brief summary and conclusions.

\section{Mathematical model}

Historically, the Klein-Gordon equation (KGE) for an electron is obtained
from the relativistic relation between the energy ${\cal E}$ and the momentum ${\bf p}$, viz.
\begin{equation}
  {\cal E}^2={\bf p}^2 c^2 + m_e^2 c^4,
\end{equation}
where $c$ is the speed of light in vacuum and $m_e$ the electron mass.
By the substitution ${\cal E}\rightarrow i\hbar \partial/\partial t$ and
${\bf p}\rightarrow -i \hbar \nabla$ in (1), where $\hbar$ is the Planck constant divided
by $2\pi$, we obtain the KGE for a free electron as
\begin{equation}
  \hbar^2 \frac{\partial^2 \psi }{\partial t^2}- \hbar^2 c^2\nabla^2\psi + m_e^2 c^4\psi=0,
\end{equation}
where $\psi$ is the electron wave function. The free-particle KGE fulfills the
continuity equation
\begin{equation}
  \frac{\partial \rho_e}{\partial t}+\nabla\cdot {\bf j}=0,
  \label{eq3}
\end{equation}
where
\begin{equation}
 \rho_e =- \frac{ie\hbar}{2 m_e c^2}\left(\psi^\ast \frac{\partial \psi}{\partial t}
-\psi\frac{\partial \psi^\ast}{\partial t}\right),
  \label{eq4}
\end{equation}
and
\begin{equation}
  {\bf j}_e= \frac{i e\hbar}{2 m_e}(\psi^\ast\nabla\psi-\psi\nabla\psi^\ast).
  \label{eq5}
\end{equation}
We have multiplied the right-hand sides of Eqs. (\ref{eq4}) and (\ref{eq5}) by the electron charge $-e$, so that
$\rho_e$ can be interpreted as the electric charge density and ${\bf j}$ as the electric current density.
Since $\rho_e$ is neither positive or negative definite, it cannot be interpreted as a probability density,
however, it can be interpreted as a charge density which need not has a definite sign.

We now wish to use the charge and current densities as sources for the self-consistent electromagnetic
scalar and vector potentials $\phi$ and ${\bf A}$ for a quantum plasma. We, therefore, let $\psi$
represent an ensemble of the electrons.  Introducing the electromagnetic potentials into the KGE,
we make the usual substitutions $i\hbar \partial/\partial t \rightarrow i\hbar \partial / \partial t + e\phi$
and $-i\hbar\nabla \rightarrow -i\hbar \nabla + e{\bf A}$, obtaining
\begin{equation}
{\cal W}^2\psi-c^2{\cal P}^2\psi-m_e^2c^4\psi=0,
\label{KG}
\end{equation}
where we have defined the energy and momentum operators as
\begin{equation}
  {\cal W}=i\hbar\frac{\partial}{\partial t}+e\phi,
\end{equation}
and
\begin{equation}
  {\cal P}=-i\hbar\nabla+e{\bf A},
\end{equation}
respectively. The electric charge and current densities are now obtained as
\begin{equation}
\rho_e=-\frac{e}{2 m_e c^2}\left[\psi^\ast {\cal W}\psi+\psi({\cal W}\psi)^\ast\right],
\label{rho_e}
\end{equation}
and
\begin{equation}
{\bf j}_e=- \frac{e}{2 m_e}\left[\psi^\ast{\cal P}\psi+\psi({\cal P}\psi)^\ast\right],
\label{j_e}
\end{equation}
respectively. We note that the charge and current densities obey the continuity equation
\begin{equation}
  \frac{\partial \rho_e}{\partial t}+\nabla\cdot {\bf j}_e=0.
  \label{continuity}
\end{equation}

The self-consistent vector and scalar potentials are obtained from the electromagnetic wave equations
\begin{equation}
\frac{\partial^2{\bf A}}{\partial t^2}+ c^2 \nabla\times(\nabla\times{\bf A})
+  \nabla\frac{\partial \phi}{\partial t}=\mu_0 c^2{\bf j}_e,
\label{vector_pot}
\end{equation}
and
\begin{equation}
   \nabla^2\phi+\nabla\cdot\frac{\partial {\bf A}}{\partial t}
=-\frac{1}{\varepsilon_0}(\rho_e+\rho_i),
   \label{scalar_pot}
\end{equation}
where $\mu_0$ is the magnetic vacuum permeability and $\varepsilon_0$ is the electric permittivity
in vacuum, and $\rho_i$ is the neutralizing positive charge density due to the ions. For immobile,
singly charged ions, one can assume that $\rho_i=e n_0$, where $n_0$ is the equilibrium ion number
density.

Using the Coulomb gauge $\nabla \cdot {\bf A}=0$, we obtain from Eqs. (\ref{vector_pot}) and (\ref{scalar_pot})
\begin{equation}
\frac{\partial^2{\bf A}}{\partial t^2}- c^2 \nabla^2{\bf A} + \nabla\frac{\partial \phi}{\partial t}
=\mu_0 c^2{\bf j}_e,
\label{wave2}
\end{equation}
and
\begin{equation}
   \nabla^2\phi=-\frac{1}{\varepsilon_0}(\rho_e+\rho_i),
   \label{Poisson}
\end{equation}
respectively. Taking the divergences of both sides of Eq.~(\ref{wave2}), we have
\begin{equation}
  \nabla^2\frac{\partial \phi}{\partial t}=\mu_0 c^2 \nabla\cdot{\bf j}_e,
\end{equation}
so that Eq. (\ref{wave2}) can be written as
\begin{equation}
\nabla^2\left(\frac{\partial^2{\bf A}}{\partial t^2}- c^2\nabla^2{\bf A}\right)
=-\mu_0 c^2 \nabla\times(\nabla\times {\bf j}_e).
\label{wave3}
\end{equation}
Equations (\ref{KG}), (\ref{Poisson}) and (\ref{wave3}) are our desired system that
describes intense laser-plasma interactions in the quantum regime.

The non-relativistic limit is obtained from Eq. (\ref{KG}) by substituting $\psi=\Psi\exp(-im_e c^2 t/\hbar)$,
and by using the condition $|\hbar\partial\Psi/\partial t|\ll m_e c^2\Psi$, together with the normalization of
$\Psi$ such that $\Psi \Psi^\ast=n_0$ is the electron number density at the equilibrium.
In this limit,  Eq. (\ref{KG}), yields the Schr\"odinger equation
\begin{equation}
  i\hbar\frac{\partial \Psi}{\partial t}+\frac{1}{2 m_e}(-i\hbar \nabla+ e{\bf A})^2\Psi +e\phi\Psi=0.
\end{equation}

Here, and in what follows, we have used a simplified model and neglected the electron degeneracy pressure.
The latter is important in dense matters where the electron degeneracy pressure appears due to the
Pauli exclusion principle.  For a non-relativistic plasma, the quantum statistical pressure has been
introduced in a nonlinear Schr\"odinger model \cite{Manfredi01}, but this has to be investigated
for relativistic quantum plasmas.

\section{Collective electrostatic oscillations and free particles}

In the absence of the electromagnetic field (viz. ${\bf A}=0$), we still have electrostatic
waves due to the charge separation between the electrons and ions. At short wavelengths, the quantum
effects become important and give rise to dispersive effects in the electrostatic wave. At these wavelengths,
there is an interplay between collective electron oscillations and free electron motion. When the wavelength
is comparable to the Compton wavelength, the electrons become relativistic, and there are relativistic
corrections to the dispersion relation for the electrostatic wave.

In the derivation of the dispersion relation for relativistic electrons, it is convenient to first make
the transformation $\psi=\widetilde{\psi}\exp(-i m_e c^2 t/\hbar)$, where the wave function $\widetilde{\psi}$
obeys the wave equation
\begin{equation}
\bigg(i \hbar\frac{\partial}{\partial t}+ m_e c^2 +e\phi\bigg)^2\widetilde{\psi}
+\hbar^2 c^2\nabla ^2\widetilde{\psi}-m_e^2c^4\widetilde{\psi}=0,
\end{equation}
and the electron charge density is
\begin{equation}
\rho_e=-\frac{i \hbar e}{2 m_e c^2}\bigg(\widetilde{\psi}^* \frac{\partial \psi}{\partial t}
-\psi\frac{\partial \widetilde{\psi}^*}{\partial t}\bigg)-\bigg(1+\frac{e\phi}{m_e
c^2}\bigg)e|\widetilde{\psi}|^2.
\end{equation}

We next linearize the system (19) by setting $\phi=\phi_1$ and $\psi=\psi_0+\psi_1$, where
$\phi_1=\widehat{\phi}\exp(i {\bf K}\cdot{\bf r}-i \Omega t)$ + complex conjugate,
$\psi_1=\widehat{\psi}_{+}\exp(i {\bf K}\cdot{\bf r}-i \Omega t)+ \widehat{\psi}_{-}
\exp(-i {\bf K}\cdot{\bf r}+i \Omega t)$, and where $|\psi_0|^2=n_0$. Separating different Fourier modes,
we obtain from (19) the dispersion relation for the electrostatic oscillations as ${\cal E}=1+\chi_e=0$,
where ${\cal E}$ is the dielectric constant and the electron susceptibility is
\begin{equation}
 \chi_e=\frac{\omega_{pe}^2[4 m_e^2 c^4-\hbar^2(\Omega^2-c^2K^2)]}
{\hbar^2(\Omega^2-c^2 K^2)^2-4 m_e^2 c^4\Omega^2},
 \label{chi_e}
\end{equation}
where $\omega_{pe}=(n_0 e^2/\varepsilon_0 m_e)^{1/2}$ is the electron plasma frequency.
We note that in the classical limit $\hbar\rightarrow 0$, we have $\chi_e=-\omega_{pe}^2/\Omega^2$,
while in the non-relativistic limit $c\rightarrow\infty$, we have $\chi_e=-\omega_{pe}^2/
(\Omega^2-\hbar^2 k^4/4 m_e^2)$.  After some reordering of terms, the dispersion relation can be
written as
\begin{equation}
  \hbar^2(\Omega^2-c^2 K^2)(\Omega^2-c^2 K^2-\omega_{pe}^2)-4m_e^2 c^4(\Omega^2-\omega_{pe}^2)=0.
  \label{ES_disp}
\end{equation}
In the classical limit $\hbar\rightarrow 0$ we obtain the Langmuir oscillations $\Omega=\omega_{pe}$, while
in the limit $c\rightarrow\infty$, we retain the  non-relativistic result
\begin{equation}
  \Omega^2=\omega_{pe}^2 + \frac{\hbar^2 k^4}{4 m_e^2}.
  \label{ES_disp2}
\end{equation}
On the other hand, in the limit $K\rightarrow 0$, Eq. (\ref{ES_disp}) yields two possibilities, one of
which is the Langmuir oscillations at the plasma frequency, $\Omega=\omega_{pe}$ and the other one is oscillations
with the frequency $\Omega=2 m_e c^2/\hbar$. The latter corresponds to a negative energy state, which can be
interpreted as positronic state.

We note that there is a non-dimensional quantum parameter
\begin{equation}
H=\hbar \omega_{pe}/m_e c^2
\label{H}
\end{equation}
in Eq. (\ref{ES_disp}) that determines the relative importance of the quantum effect. Typical values
are $H=10^{-4}$ for the electron number density $n_e \sim 10^{30}\,\mathrm{m}^{-3}$ in solid density
laser-plasma experiments and $H=0.007$ may be representable of modern laser-compressed matter experiments
with $n_e\sim 10^{34}\,\mathrm{m}^{-3}$.  This corresponds to $\omega_{pe}=8\times 10^{16}\,\mathrm{s^{-1}}$
and $\lambda_e=4\times 10^{-9}\,\mathrm{m}$ for $H=10^{-4}$, and $\omega_{pe}=5.4\times 10^{18}\,\mathrm{s}^{-1}$
and $\lambda_e=5.5\times 10^{-11}\,\mathrm{m}$ for $H=0.007$, where $\lambda_e=c/\omega_{pe}$ is the electron
skin depth.  The non-relativistic result (\ref{ES_disp2}) is valid for electrostatic waves with wave numbers in
the range $1<K \lambda_e <1/H$.  For $K \lambda_e <1$, the quantum corrections to $\omega\approx \omega_{pe}$
are different from (\ref{ES_disp2}) and turns the wave frequency slightly lower than $\omega_{pe}$.  However,
this effect is negligible for small values of $H$, and may be smaller than the degeneracy electron pressure
effect, which is neglected here. On the other hand, the limit $K \lambda_e >1/H$ corresponds to relativistic
particles with $K > 1/\lambda_C$, where $\lambda_C=\hbar/m_e c\approx 3.9\times 10^{-13}\,\mathrm{m}$ is
the reduced Compton wavelength.  For $\omega_{pe} \rightarrow 0$, we obtain the relativistic free particle
dispersion relation
\begin{equation}
  \Omega=\mp\frac{m_e c^2}{\hbar}+ \sqrt{\frac{m_e^2 c^4}{\hbar^2}+ c^2 K^2},
\end{equation}
where the upper sign (-) corresponds to the motion of a free electron and the lower sign (+) can be interpreted
as the motion of a free positron.

\begin{figure}
\includegraphics[width=8.5cm]{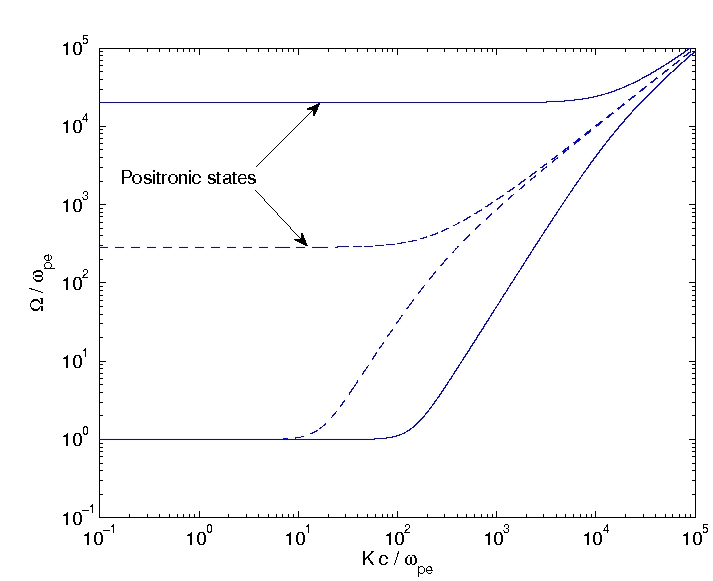}
\caption{Dispersion curves ($\Omega$ vs. $K$) for the electrostatic oscillations for $H=10^{-4}$ (solid curves),
$H=0.007$ (dashed curves), where $H=\hbar\omega_{pe}/m_e c^2$. For $K \lambda_e >1/H$, the particle motion
turns from weakly relativistic to ultra-relativistic.}
\end{figure}

In Fig. 1, we have plotted the solutions of the dispersion relation (\ref{ES_disp}) for $H=10^{-4}$
and $H=0.007$. Both the electron plasma oscillations and the positronic states are shown. The electron
plasma oscillations have a cutoff frequency $\omega\rightarrow\omega_{pe}$ when $K\rightarrow0$,
while the positronic states have a cutoff frequency $\omega\rightarrow 2 m_e c^2/\hbar$,
corresponding to $\omega/\omega_{pe}\rightarrow 1/H$ at $K\rightarrow 0$ in Fig. 1. For the electron plasma
oscillations, the increase in the wave frequency due to the quantum effect becomes noticeable
approximately where $K\lambda_e = 1/\sqrt{H}$, or $K \approx (4\pi n_0/a_B)^{1/4}$ where
$a_B= 4\pi \varepsilon_0\hbar^2/m_e e^2 \approx 5.3\times 10^{-11}\,\mathrm{m}$ is the Bohr radius.
This corresponds to a wavelength of $2\pi/K\approx 2.8\times 10^{-10}\,\mathrm{m}$ for $H=10^{-4}$ and
$2\pi/K\approx 5\times 10^{-11}\,\mathrm{m}$ for $H=0.007$.

The positronic states are associated with, for example, the Zitterbewegung effect \cite{Fuda82,Gerritsma10},
in which the interference between the positive and negative energy states are predicted to give oscillations
on Compton wavelength scales in space. The Zitterbewegung effect is still debated and has not yet been observed
in experiments.

\section{Nonlinear electromagnetic wave propagation and self-induced transparency}

It is well-known \cite{Akhiezer56} that a large amplitude electromagnetic wave propagating
in a classical plasma changes the dispersive properties of the plasma due to the relativistic
mass increase of the electrons. We show here that the same effect occurs in our Klein-Gordon-Maxwell
system.

We consider for simplicity the propagation of a right-hand circularly polarized electromagnetic (CPEM) wave
of the form ${\bf A}=A_0[\widehat{\bf x}\cos(k_0 z-\omega_0 t)-\widehat{\bf y}\sin(k_0 z-\omega_0 t)]$,
where $\omega_0$ is the wave frequency and $k_0$ the wavenumber. Due to the circular polarization,
the oscillatory parts in the nonlinear term proportional to $A^2$ in the Klein-Gordon equation vanish.
Assuming that $\psi$ depends only on time and not on space, and that $\phi=0$, we obtain from
Eq. (\ref{KG})
\begin{equation}
  \hbar^2 \frac{\partial^2\psi }{\partial t^2} +m_e^2 c^4 \gamma_A^2\psi= 0,
  \label{KG_time}
\end{equation}
where $\gamma_A=\sqrt{1+e^2 A_0^2/m_e^2 c^2}$ can be interpreted as the relativistic gamma
factor due to the electron mass increase in the CPEM wave field. Equation (\ref{KG_time}) has
the solution
\begin{equation}
  \psi=\psi_0\exp(-i m_e c^2 \gamma_A t/\hbar),
  \label{psi}
\end{equation}
where the constant $\psi_0$ is determined by assuming the constant density $\rho_e=-e n_0$ in Eq. (\ref{rho_e}).

Inserting (\ref{psi}) into (\ref{rho_e}) with $\rho_e=-e n_0$, we obtain
\begin{equation}
  |\psi_0|^2=\frac{n_0}{\gamma_A}.
\end{equation}
On the other hand, inserting (\ref{psi}) into (\ref{j_e}) we have
\begin{equation}
  {\bf j}_e=-\frac{e^2 |\psi_0|^2}{m_e} {\bf A}=-\frac{e^2 n_0}{\gamma_A m_e}{\bf A},
\end{equation}
which can be inserted into (\ref{wave2}) to obtain
\begin{equation}
\frac{1}{c^2}\frac{\partial^2{\bf A}}{\partial t^2}-\nabla^2{\bf A}
=-\frac{\mu_0 e^2 n_0}{\gamma_A m_e}{\bf A}.
\label{plane_wave}
\end{equation}
Equation (\ref{plane_wave}) admits the nonlinear dispersion relation
\begin{equation}
  \omega_0^2=c^2 k^2+\frac{\omega_{pe}^2}{\gamma_A},
  \label{omega0}
\end{equation}
which predicts a relativistic downshift of the CPEM wave frequency due to the
relativistic electron mass increase in the CPEM wave field. Since the effective plasma
frequency is decreased by a factor $1/\sqrt{\gamma_A}$, the model predicts the well-known
self-induced transparency where the CPEM wave can propagate at frequencies below the
electron plasma frequency. This is identical to the case of classical plasmas \cite{Akhiezer56}.

\section{Stimulated Raman scattering and modulational instabilities}

We now consider the instability of an intense CPEM wave in the quantum regime. In the presence of intense
electromagnetic waves, we have the relativistic down-shift in the wave frequency given in (\ref{omega0}),
 as well as the possibility of exciting electrostatic oscillations via the parametric instabilities.
As an example, we will here consider stimulated Raman scattering instability, in which an
intense electromagnetic wave decays into a daughter EM wave and an electron plasma wave.
The two-plasmon decay instability, in which the CPEM wave decays into two electrostatic waves,
 will be treated elsewhere.

It is convenient to first introduce the transformation $\psi=\widetilde{\psi}\exp(-i \gamma_A m_e c^2 t/\hbar)$,
where $\gamma_A=\sqrt{1+e^2 A_0^2/m_e^2 c^2}$ and $A_0$ is the amplitude of the EM carrier wave ${\bf A}_0$.
The wavefunction $\widetilde{\psi}$ obeys the modified Klein-Gordon equation
\begin{equation}
  \bigg(i \hbar\frac{\partial}{\partial t}+ \gamma_A m_e c^2 +e\phi\bigg)^2\widetilde{\psi}
  -c^2(-i\hbar\nabla+e{\bf A})^2\widetilde{\psi}-m_e^2c^4\widetilde{\psi}=0,
  \label{KG5}
\end{equation}
and the electron charge density is given by
\begin{equation}
  \rho_e=-\frac{i \hbar e}{2 m_e c^2}\bigg(\widetilde{\psi}^* \frac{\partial \psi}{\partial t}
-\psi\frac{\partial \widetilde{\psi}^*}{\partial t}\bigg)
  -\bigg(\gamma_A+\frac{e\phi}{m_e c^2}\bigg)e|\widetilde{\psi}|^2.
\end{equation}

Now, we linearize our system by introducing $\widetilde{\psi}({\bf r},t)
=\widetilde{\psi}_0+\widetilde{\psi}_1({\bf r},t)$
(where $\widetilde{\psi}_0$ is assumed to be constant),
${\bf A}= {\bf A}_0({\bf r},t)+{\bf A}_1({\bf r},t)$, and $\phi({\bf r},t)=\phi_1({\bf r},t)$.
Using $\rho_i=e n_0$ into Eq. (\ref{Poisson}), we note that the equilibrium quasi-neutrality requires
that $\widetilde{\psi}_0$ is normalized such that $|\widetilde{\psi}_0|^2=n_0/\gamma_A$.
Using that ${\bf A}_0$ fulfills the plane wave equation (\ref{plane_wave}), the
linearized KGE (\ref{KG5}), Poisson's equation (\ref{Poisson}) and
the EM wave equation (\ref{wave3}) then become
\begin{equation}
  \begin{split}
  &\hbar^2 \bigg(-\frac{\partial^2 \widetilde{\psi}_1}{\partial t^2} +c^2\nabla^2 \widetilde{\psi}_1\bigg)
   +2 i\hbar \gamma_A m_e c^2 \frac{\partial \widetilde{\psi}}{\partial t}
  + 2i \hbar c^2 e {\bf A}_0\cdot \nabla \widetilde{\psi}_1
\\
  &+\bigg(2\gamma_A m_e c^2e\phi_1+i\hbar e \frac{\partial \phi_1}{\partial t}\bigg)\widetilde{\psi}_0
  -2 c^2 e^2 {\bf A}_0\cdot{\bf A}_1\widetilde{\psi}_0=0,
  \end{split}
  \label{KG_linear}
\end{equation}
\begin{equation}
  \begin{split}
   &\nabla^2\phi_1=\frac{ie\hbar}{2\varepsilon_0 m_e c^2}\bigg(\widetilde{\psi}_0^\ast
\frac{\partial \widetilde{\psi}_1}{\partial t}
    -\widetilde{\psi}_0\frac{\partial \widetilde{\psi}_1^\ast}{\partial t}\bigg)
+\frac{e\gamma_A}{\varepsilon_0}( \widetilde{\psi}_0^\ast \widetilde{\psi}_1
+ \widetilde{\psi}_0 \widetilde{\psi}_1^\ast)
    +\frac{\omega_{pe}^2}{\gamma_A c^2}\phi_1,
  \end{split}
\end{equation}
and
\begin{equation}
  \begin{split}
  &\nabla^2\bigg(
    \frac{\partial^2{\bf A}_1}{\partial t^2}-c^2\nabla^2{\bf A}_1+\frac{\omega_{pe}^2}{\gamma_A}{\bf A}_1
  \bigg)
=\frac{\omega_{pe}^2}{n_0}\nabla\times\{\nabla\times[{\bf A}_0(\widetilde{\psi}_0^\ast\widetilde{\psi}_1
+\widetilde{\psi}_0\widetilde{\psi}_1^\ast)]\},
  \end{split}
\end{equation}
respectively. We note that the term proportional to ${\bf A}_0\cdot \nabla \widetilde{\psi}_1$ in
Eq. (\ref{KG_linear}) gives rise to the two-plasmon decay, which we, however, do not consider here.

We now introduce the Fourier representations $\widetilde{\psi}=\widehat{\psi}_{+}\exp(-i\Omega t+i{\bf K}\cdot {\bf r})
+\widehat{\psi}_{-}\exp(i\Omega t-i{\bf K}\cdot {\bf r})$,
$\phi_1=\widehat{\phi} \exp(-i\Omega t+i{\bf K}\cdot {\bf r})$+ c.c.,
${\bf A}_0=(1/2)\widehat{\bf A}_0\exp(-\omega_0 t+{\bf k}_0\cdot{\bf r})$+ c.c., and
${\bf A}_1=[\widehat{\bf A}_{+}\exp(-i\omega_{+}t+i{\bf k}_{+}\cdot{\bf r})
+\widehat{\bf A}_{-}\exp(-i\omega_{-}t+i{\bf k}_{-}\cdot{\bf r})]$ +
c.c., where we introduced $\omega_{\pm}=\omega_0\pm \Omega$ and ${\bf k}_{\pm}={\bf k}_0\pm{\bf K}$,
and c.c.~stands for complex conjugate. In one of the steps, we take the scalar product of both sides
of the EM wave equation by $\widehat{\bf A}_0^\ast$ and use the fact that $\widehat{\bf A}_0^\ast\cdot
[{\bf k}_{\pm}\times({\bf k}_\pm\times \widehat {\bf A}_0)] =({\bf k}_\pm\times\widehat{\bf A}_0)\cdot
(\widehat{\bf A}_0^\ast \times{\bf k}_\pm)=-|{\bf k}_{\pm}\times \widehat {\bf A}_0|^2$.
Separating different Fourier modes and eliminating the Fourier coefficients, we find the nonlinear
dispersion relation
\begin{equation}
\begin{split}
&1+\frac{1}{\widetilde{\chi}_e}=\frac{(c^2 K^2-\Omega^2+\omega_{pe}^2/\gamma_A)}
{[4\gamma_A^2 m_e^2 c^4-\hbar^2(\Omega^2-c^2 K^2)]}
 \bigg[\frac{c^2 e^2|{\bf k}_+\times \widehat{\bf A}_0|^2}{k_{+}^2 D_A(\omega_{+},{\bf k}_{+})}
+\frac{c^2 e^2|{\bf k}_{-}\times \widehat{\bf A}_0|^2}{k_{-}^2 D_A(\omega_{-},{\bf k}_{-})}\bigg],
\end{split}
\label{nonlin_disp2}
\end{equation}
where the electromagnetic sidebands are governed by
$D_A(\omega_\pm,{\bf k}_\pm)=c^2 k_\pm^2-\omega_\pm^2+\omega_{pe}^2/\gamma_A$.
The electric susceptibility in the presence of the laser field is given by
\begin{equation}
  \widetilde{\chi}_e=\frac{\omega_{pe}^2[4\gamma_A^2 m_e^2 c^4-\hbar^2(\Omega^2-c^2 K^2)]}
{\gamma_A[\hbar^2(\Omega^2-c^2K^2)^2-4\gamma_A^2 m_e^2 c^4 \Omega^2]}.
\end{equation}
After reordering of terms, the nonlinear dispersion relation (\ref{nonlin_disp2}) can be written
as
\begin{equation}
\begin{split}
&1-\frac{\omega_{pe}^2}{4 \gamma_A^3 m_e^2 c^2}\frac{(c^2 K^2-\Omega^2+\omega_{pe}^2/\gamma_A)}
{\widetilde{D}_L(\Omega,{\bf K})}
\bigg[\frac{e^2|{\bf k}_+\times\widehat{\bf A}_0|^2}{k_{+}^2 D_A(\omega_{+},{\bf k}_{+})}
+\frac{e^2|{\bf k}_{-}\times\widehat{\bf A}_0|^2}{k_{-}^2 D_A(\omega_{-},{\bf k}_{-})}\bigg]=0,
\end{split}
\label{nonlin_disp}
\end{equation}
where the electron plasma oscillations in the presence of the laser field are represented by
\begin{equation}
\begin{split}
  &\widetilde{D}_L(\Omega,{\bf K})=
\frac{\omega_{pe}^2}{\gamma_A}-\Omega^2+\frac{\hbar^2 (\Omega^2-c^2 K^2)}
{4\gamma_A^2 m_e^2 c^4}\bigg(\Omega^2-c^2K^2-\frac{\omega_{pe}^2}{\gamma_A}\bigg).
\end{split}
\end{equation}
We note that $D_L=0$ gives the dispersion relation for pure electrostatic oscillations in the presence of
a large amplitude electromagnetic wave.

In the classical limit $\hbar \rightarrow 0$, we have
$\widetilde{\chi}_e=-\omega_{pe}^2/\Omega^2\gamma_A$, and the nonlinear dispersion relation
takes the form
\begin{equation}
\begin{split}
&1-\frac{\Omega^2\gamma_A}{\omega_{pe}^2}=\frac{(c^2 K^2-\Omega^2+\omega_{pe}^2/\gamma_A)}{4 \gamma_A^2 m_e^2 c^2}
 \bigg[\frac{e^2|{\bf k}_+\times\widehat {\bf A}_0|^2}{k_{+}^2 D_A(\omega_{+},{\bf k}_{+})}
+\frac{e^2|{\bf k}_{-}\times\widehat{\bf A}_0|^2}{k_{-}^2 D_A(\omega_{-},{\bf k}_{-}) }\bigg],
\end{split}
\end{equation}
which can be written in a more familiar form as
\begin{equation}
\begin{split}
&1-\left(\frac{c^2 K^2}{D_L}+1\right)\frac{\omega_{pe}^2}{4 \gamma_A^3 m_e^2 c^2}
 \bigg[\frac{e^2|{\bf k}_+\times\widehat{\bf A}_0|^2}{k_{+}^2 D_A(\omega_{+},{\bf k}_{+})}
+\frac{e^2|{\bf k}_{-}\times\widehat{\bf A}_0|^2}{k_{-}^2 D_A(\omega_{-},{\bf k}_{-}) }\bigg]=0,
\end{split}
\end{equation}
with $D_L=\omega_{pe}^2/\gamma_A-\Omega^2$. These results can be compared with, for example, the
dispersion relations obtained in Refs. \cite{Sakharov94,Guerin95,Quesnel97} for the relativistic case and
in \cite{Drake74} for the non-relativistic case.

To proceed with the numerical evaluation of the nonlinear dispersion relation, we
choose a coordinate system such that the CPEM takes the form
$\widehat{\bf A}_0=(\widehat{\bf x}+i\widehat{\bf y})\widehat{A}_0$ and
${\bf k}_0=k_0\widehat{\bf z}$, and, without loss of generality, we choose
${\bf K}=K_{||}\widehat{\bf z}+K_\perp\widehat{\bf y}$.
Then, we have $K^2=K_{||}^2+K_\perp^2$, $\gamma_A=\sqrt{1+e^2 |\widehat{A}_0|^2/m_e^2 c^2}$,
$|{\bf k}_\pm\times \widehat{\bf A}_0|^2=[2(k_0 \pm K_{||})^2+K_\perp^2]|\widehat{A}_0|^2$,
and $k_\pm^2=(k_0\pm K_{||})^2+K_\perp^2$. We also use that the carrier wave ${\bf A}_0$ obeys the nonlinear
dispersion relation $\omega_0=\sqrt{c^2 k_0^2+\omega_{pe}^2/\gamma_A}$.

\begin{figure}[htb]
\includegraphics[width=8.5cm]{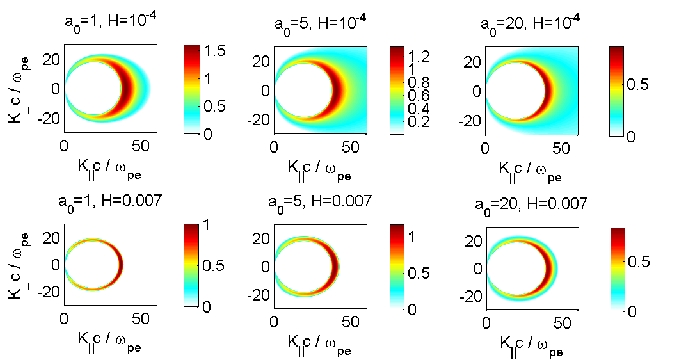}
\caption{Growth rate ($\Omega_I/\omega_{pe}$ vs. $K_{||}$ and $K_\perp$) for
stimulated Raman scattering in the presence of a large amplitude CPEM wave,
for the amplitudes $a_0=1$, $a_0=5$, and $a_0=20$ (left to right panels) where $a_0=e |\widehat{A}_0|/m_e c$
for $H=10^{-4}$ (top panels) and $H=0.007$ (bottom panels).
}
\end{figure}

We now assume that the wave frequency is complex valued, $\Omega=\Omega_R+i\Omega_I$, where
$\Omega_R$ is the real frequency and $\Omega_I$ the growth rate, and solve numerically the
dispersion relation (\ref{nonlin_disp}) for $\Omega$.  In Fig. 2, we have plotted the growth rate
for stimulated Raman scattering instability as a function of the wavenumbers $K_{||}$ and $K_\perp$,
for a few values of $a_0=e|\widehat{A}_0|/m_e c$ and $H=\hbar \omega_{pe}/m_e c^2$. For all cases in Fig. 2,
we used $k_0 c /\omega_{pe}=20$, which corresponds to a wavelength of $1.25\times 10^{-9}\,\mathrm{m}$
for $H=10^{-4}$ and to $1.7\times 10^{-11}\,\mathrm{m}$ for $H=0.007$, which is in the X-ray regime.
We observe that for $H=10^{-4}$, there is a broad spectrum of unstable waves, in particular for $a_0=5$
and $a_0=10$. For $H=0.007$, we observe a reduction in the spectrum of unstable waves and the
growth rate (relative to the electron plasma frequency) is slightly reduced.
This is due to the fact that the wavelength of the unstable electrostatic oscillation approaches the
critical wavelength, where quantum dispersive effects become important compared to
the plasma frequency oscillations. For $H=10^{-4}$, this wavelength is
$\lambda_{crit}=2\pi/(4 \pi n_0/a_B)^{1/4}\approx 5\times 10^{-10}\,\mathrm{m}$,
corresponding to a critical wavenumber $K_{crit}=1.25\times 10^{10}\,\mathrm{m^{-1}}$, and
for $H=0.007$, we have $\lambda_{crit}=2\pi/(4 \pi n_0 a_B)\approx 2.8\times 10^{-11}\,\mathrm{m}$,
corresponding to $K_{crit}=2.25\times 10^{11}\,\mathrm{m^{-1}}$. Hence, for $H=0.007$ we have
$k_0\approx K_{crit}$, which leads to the reduction of the growth rate due to the quantum dispersion effect.

Furthermore, it should be mentioned that we do not find Raman-type instabilities involving the
positronic states in Fig. 1.  This is consistent from the point of view of the conservation of
charges, since the production of positrons would violate the conservation of electric charges.

\begin{figure}[htb]
\includegraphics[width=8.5cm]{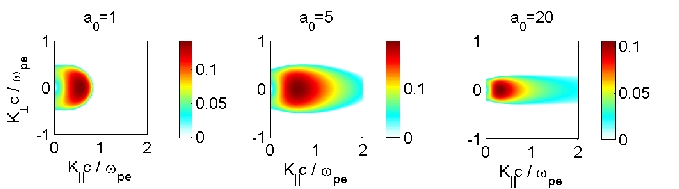}
\caption{The growth rate ($\Omega_I/\omega_{pe}$ vs. $K_{||}$ and $K_\perp$) for the modulational instability
in the presence of a large amplitude CPEM wave, for the amplitudes $a_0=1$ (left panel), $a_0=5$ (middle panel)
and $a_0=20$ (right panel) where $a_0=e |\widehat{A}_0|/m_e c$. We used $H=0.007$ for all cases.
}
\end{figure}

In addition to stimulated Raman scattering instabilities, we have also the modulational instability
that dominates for the  pump frequencies $\omega_0 < 2\omega_{pe}/\sqrt{\gamma_A}$, and the corresponding
wavenumbers $k_0< (\omega_{pe}/c) \sqrt{3/\gamma_A}$. The modulational instability usually occurs for
small modulation wavenumbers, and saturates nonlinearly by the formation of relatively small localized
structures/solitary waves. In the past, such nonlinear structures have been studied for
classical plasmas in 1D \cite{Marburger75} and 3D \cite{Gersten75}. We have investigated the
modulational instability for the CPEM dipole field with $k_0=0$, and have plotted the results
in Fig. 3 for different amplitudes $a_0=1$, $a_0=5$ and $a_0=20$. We find that the growth rate
is relatively insensitive to the quantum parameter $H$. We have used $H=0.007$ in Fig. 3, but $H=0$ gives almost
identical results. This is understandable since the modulational instability takes place on relatively
large scales and the quantum effect is thus negligible. However, we will investigate
the quantum effect on the relatively small scale nonlinear structures below.

\section{Relativistic optical solitary waves}

Here we illustrate the existence of large amplitude localized CPEM wave excitations
at the quantum scale in our system. We restrict our investigation to one-space dimension,
which has also been studied for classical plasmas \cite{Marburger75}.

Far away from the local excitation, one can assume that the dynamics of the plasma is non-relativistic.
To shorten the algebraic steps, it is convenient first to introduce a new wave-function $\Psi(z,t)$
and the potential $\Phi$ via the transformations $\psi({\bf r},t)=\Psi(z,t)\exp(-i m_e c^2 t/\hbar)$
and $\phi=\Phi-{m_e c^2}/{e}$, and which satisfy the KGE
\begin{equation}
\begin{split}
&\left(i\hbar\frac{\partial}{\partial t}+e\Phi\right)^2\Psi+\hbar^2 c^2 \frac{\partial^2\Psi}{\partial z^2}
-\gamma_A^2 m_e^2 c^4\Psi=0,
\end{split}
\label{KG3}
\end{equation}
where $\gamma_A=(1+{e^2 A^2}/{m_e^2 c^2})^{1/2}$. In this gauge, the wave function $\Psi$ is non-oscillatory
in time, and the new potential takes the value $\Phi=e/m_e c^2$, far away from the solitary wave
where the plasma is at rest.  The electron charge density is expressed as
\begin{equation}
  \rho_e=-\frac{e}{m_e c^2}\left[
  \frac{i\hbar}{2}\left(
    \Psi^\ast\frac{\partial \Psi}{\partial t}-\Psi\frac{\partial \Psi^\ast}{\partial t}
  \right)
  +e\Phi|\Psi|^2
  \right].
  \label{e-charge}
\end{equation}

We now study quasi-steady state structures propagating with a constant speed $v_0$, so that
$\phi=\phi(\xi)$ and $A^2=A^2(\xi)$, where $\xi=z-v_0 t$ and $A^2=|{\bf A}|^2$.
The CPEM wave vector potential is of the form ${\bf A}=A(\xi)[\widehat{\bf x}\cos(k_0 z-\omega_0 t)
-\widehat{\bf y}\sin(k_0 z-\omega_0 t)]$. It is convenient
to introduce the eikonal representation $\Psi=P(\xi)\exp[i\theta(\xi)]$,
where $P$ and $\theta$ are real-valued.  Then, the KGE (\ref{KG3}) takes the form
\begin{equation}
\begin{split}
  & \hbar^2(c^2-v_0^2)\left[
    \frac{d^2 P}{d\xi^2}-P\left(\frac{d \theta}{d\xi}\right)^2+2 i \frac{d P}{d \xi}\frac{d\theta}{d\xi}
    +i P \frac{d^2\theta}{d\xi^2}
  \right]
-i \hbar e v_0 \frac{d\Phi }{d\xi} P
\\
 & -2i\hbar e \Phi v_0 \left(\frac{dP}{d\xi}+ i P\frac{d\theta}{d\xi}\right)
+(e^2\Phi^2-m_e^2 c^4\gamma_A^2)P=0.
\end{split}
\label{KG4}
\end{equation}
Setting the imaginary part of Eq.~(\ref{KG4}) to zero, we obtain
\begin{equation}
  2 U \frac{dP}{d\xi}+ P\frac{d U}{d\xi}=0,
\label{PU}
\end{equation}
where
\begin{equation}
  U=\hbar^2(c^2-v_0^2)\frac{d\theta}{d\xi}-\hbar e\Phi v_0.
\end{equation}
The solution of Eq. (\ref{PU}) is $P^2 U=D=$ constant. Using the boundary conditions
$P^2=n_0$, $\phi=0$ (hence $\Phi=m_e c^2/e$) and $d/d\xi=0$ at $|\xi|=\infty$, we have $D=-n_0 \hbar m_e c^2
v_0$. Hence, we obtain
\begin{equation}
  \frac{d\theta}{d\xi}=\frac{v_0 m_e \gamma_0^2}{\hbar}\left(
  \frac{e\Phi}{m_e c^2}-\frac{n_0}{P^2} \right),
  \label{theta}
\end{equation}
where we have denoted
\begin{equation}
  \gamma_0=\frac{1}{\sqrt{1-v_0^2/c^2}}.
\end{equation}
The electron charge density (\ref{e-charge}) now takes the form
\begin{equation}
  \rho_e=-\frac{e}{m_e c^2}\left(\hbar v_0 \frac{d\theta}{d\xi}+e\Phi\right)P^2,
  \label{rho_e4}
\end{equation}
which, by using Eq. (\ref{theta}), can be written as
\begin{equation}
  \rho_e=-e n_0 \gamma_0^2\left(
  -\frac{v_0^2}{c^2}+ \frac{e\Phi}{m_e c^2}\frac{P^2}{n_0} \right),
  \label{rho_e2}
\end{equation}
and hence Poisson's equation (\ref{Poisson}), with $\rho_i=e n_0$, becomes
\begin{equation}
  \frac{d^2\Phi}{d \xi^2}=\frac{e n_0\gamma_0^2 }{\varepsilon_0}
  \left(
  \frac{e\Phi}{m_e c^2}\frac{P^2}{n_0}-1 \right).
  \label{Poisson2}
\end{equation}
On the other hand, by setting the real part of Eq. (\ref{KG4}) to zero, we have
\begin{equation}
\begin{split}
  &\frac{\hbar^2 c^2}{\gamma_0^2}\left[\frac{d^2P}{d\xi^2}-P\left(\frac{d\theta}{d\xi}\right)^2\right]
  +2\hbar e\Phi v_0 P\frac{d\theta}{d\xi}+( e^2 \Phi^2-m_e^2 c^4\gamma_A^2)P=0,
\end{split}
\end{equation}
which, by using Eq. (\ref{theta}), can be written as
\begin{equation}
  \hbar^2\frac{d^2 P}{d\xi^2} +m_e^2 c^2 \gamma_0^4 \bigg[\frac{e^2\Phi^2}{m_e^2 c^4}
  -\frac{v_0^2}{c^2}\frac{n_0^2}{P^4}-\frac{\gamma_A^2}{\gamma_0^2} \bigg]P=0.
  \label{P}
\end{equation}

Finally, inserting the ansatz ${\bf A}=A(\xi)[\widehat{\bf x}\cos(k_0 z-\omega_0 t)
-\widehat{\bf y}\sin(k_0 z-\omega_0 t)]$ for the vector potential, together with the current
\begin{equation}
  {\bf j}_e=-\frac{e^2}{m_e}|\psi|^2 {\bf A}=-\frac{e^2}{m_e}P^2 {\bf A},
\end{equation}
into Eq. (\ref{wave3}), we obtain the EM wave equation
\begin{equation}
  \frac{d^2 A}{d \xi^2}+\frac{\omega_{pe}^2}{c^2}\bigg[\lambda +\gamma_0^2\bigg(1-\frac{P^2}{n_0}\bigg)\bigg]A=0,
  \label{A}
\end{equation}
where we used $k_0=\omega_0 v_0/c^2$, and where $\lambda=(c^2/\omega_{pe}^2)
(\omega_0^2-\omega_p^2-c^2k_0^2)/(c^2-v_0^2)=\omega_0^2/\omega_{pe}^2-\gamma_0^2$
is a nonlinear eigenvalue of the system that determines the wave frequency $\omega_0$.

The coupled system (\ref{Poisson2}), (\ref{P}) and (\ref{A}) describes the profile of electromagnetic
solitary waves in a quantum plasma. It has the conserved quantity ${\cal H}=0$, where
\begin{equation}
  \begin{split}
  &{\cal H}=-\frac{c^2}{\omega_{pe}^2}\left(\frac{d}{d\xi}\frac{e\Phi}{m_e c^2}\right)^2
            +\frac{\hbar^2}{\gamma_0^2 m_e^2 c^2 n_0}\left(\frac{dP}{d\xi}\right)^2
  +\frac{c^2}{\omega_{pe}^2\gamma_0^2}\left(\frac{d}{d\xi}\frac{e A}{m_e c}\right)^2
   +\left(\frac{\lambda}{\gamma_0^2}+1-\frac{P^2}{n_0}\right)\frac{e^2 A^2}{m_e^2 c^2}
  \\
  &+\gamma_0^2\frac{e^2\Phi^2}{m_e^2 c^4}\frac{P^2}{n_0}-2\gamma_0^2\frac{e\Phi}{m_e c^2}
+\frac{v_0^2\gamma_0^2}{c^2}\left(\frac{n_0}{P^2}-1\right)
  -\frac{P^2}{n_0}+1+\gamma_0^2.
  \end{split}
\end{equation}
The conserved quantity ${\cal H}$ can be used to check that the numerical scheme used to solve
the nonlinear system (\ref{Poisson2}), (\ref{P}) and (\ref{A}) produces correct results.

\begin{figure}[htb]
\includegraphics[width=8.5cm]{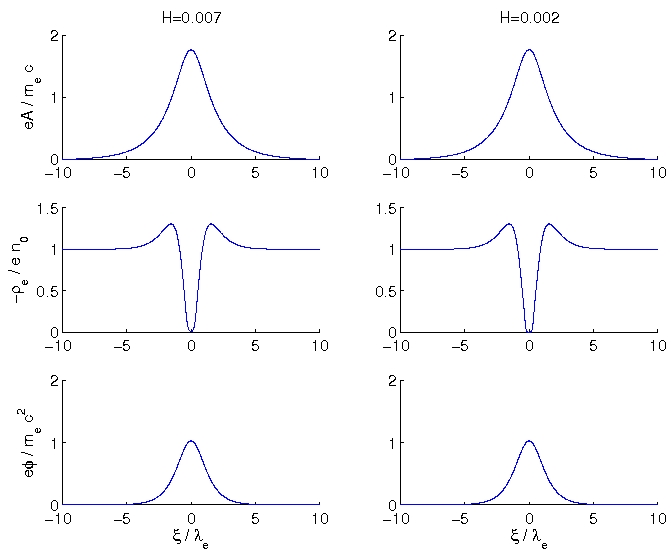}
\caption{ The spatial profiles of the vector potential, the electron charge density and the
electrostatic potential (top to bottom panels), for $H=0.007$ (left panels) and $H=0.002$ (right panels)
for standing solitary waves ($v_0=0$), and for $\lambda=-3.4$.
}
\label{fig:comparison3}
\end{figure}

\begin{figure}[htb]
\includegraphics[width=4.5cm]{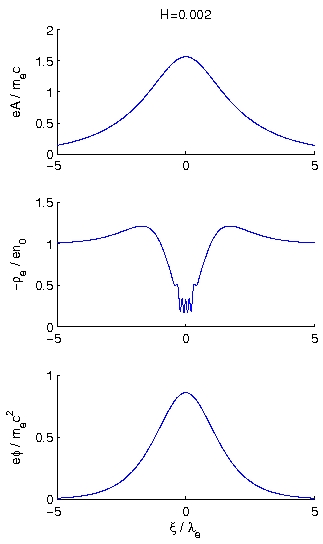}
\caption{ The spatial profiles of the vector potential, the electron charge density and the electrostatic
potential (top to bottom panels), for $H=0.002$, $v_0=0.0059 c$, and $\lambda=-3.4$.
}
\label{fig:comparison4}
\end{figure}

In Fig. \ref{fig:comparison3}, we have compared the present model with our previous results
in Ref. \cite{Shukla07} where we used a simplified model to describe the nonlinear interaction
interaction of intense CPEM wave with a quantum plasma. We used the same parameters as in Fig. 2,
of Ref. \cite{Shukla07} to produce the profiles of the CPEM wave potential, the electron charge density
and the electrostatic potential. We observe that the present results are almost identical to
our previous work \cite{Shukla07}. For our sets of parameters, the quantum effect on the
profiles of the solitary waves are small, and there is only a slight difference in the profiles
of the electron density for the two values $H=0.002$ and $H=0.007$. For standing solitary waves, such
as the ones in Fig. \ref{fig:comparison3}, the solutions are localized with exponentially decaying tails.
By linearizing the system (\ref{Poisson2}), (\ref{P}) and (\ref{A}) one can show that far away from the
soliton, $A$ decays as $\exp(\sqrt{-\lambda}\xi)$, while $P$ and $\phi$ are proportional to $\exp(K \xi)$,
where $K$ is found from the dispersion relation
\begin{equation}
  \hbar c^2 K^2 (c^2 K^2+\gamma_0\omega_{pe}^2)-4\gamma_0^4m_e^2 c^4 (v_0^2 K^2-\omega_{pe}^2)=0.
  \label{Disp_K}
\end{equation}
For $v_0=0$ (and $\gamma_0=1$), we see immediately that there exist only complex-valued $K$, which
means that the quasi-stationary wave solutions decay exponentially far away from the solitary wave.
However, in the classical limit $\hbar\rightarrow 0$, we instead have the plasma wake oscillations
given by real valued $K=\omega_{pe}/v_0$. Hence, an electromagnetic pulse will create an oscillatory
wake that extends far away from the EM pulse. In one-space dimension, there also exist special
classes of propagating localized EM envelope solutions \cite{Kaw92,Saxena06}.
In addition to the wake oscillations, we also have quantum oscillations in quantum plasmas.
In Fig. 5, we show an example of a slowly moving envelope soliton, where small-scale oscillations
in the charge density are clearly visible.

We note that the cold fluid results can be retained in the classical limit $\hbar\rightarrow 0$.
Then, Eq.~(\ref{P}) can be written as
\begin{equation}
  \frac{e^2\Phi^2}{m_e^2 c^4}=\frac{v_0^2}{c^2}\frac{n_0^2}{P^4}+\frac{\gamma_A^2}{\gamma_0^2}.
  \label{P2}
\end{equation}
By setting $\rho_e=-e n_e$, where $n_e$ is the electron number density, in Eq. (\ref{rho_e2}) and solving for $P$,
we obtain
\begin{equation}
  \frac{P^2}{n_0}=\bigg[
    1+\frac{1}{\gamma_0^2}\bigg(
    \frac{n_e}{n_0}-1
    \bigg)
  \bigg]\frac{m_e c^2}{e\Phi},
\end{equation}
which can be inserted into (\ref{P2}) to obtain
\begin{equation}
  \bigg[
  \frac{1}{\gamma_0^2}\frac{n_e}{n_0}
  +\frac{v_0^2}{c^2}
  \bigg]/\bigg[
  \frac{n_e^2}{n_0^2}-\frac{v_0^2}{c^2}\bigg(\frac{n_e}{n_0}-1\bigg)^2
  \bigg]^{1/2}
  =\frac{e\Phi}{\gamma_A m_e c^2},
  \label{fluid}
\end{equation}
which relates $n_e$ to $\Phi$ and $\gamma_A$ at a given speed $v_0$. The relation (\ref{fluid}) can
also be obtained from the cold electron fluid model \cite{Kaw92} and hence confirms the classical limit of the
quantum model used here.  If, furthermore, we assume standing waves such that $v_0=0$, then we have
from (\ref{fluid})
\begin{equation}
  \frac{e\Phi}{m_e c^2}=\gamma_A.
\end{equation}
Solving for $\Phi$ and inserting the result into Poisson's equation (\ref{Poisson2}), we have
\begin{equation}
  \lambda_e^2 \frac{d^2\gamma_A}{d \xi^2}=
  \gamma_A \frac{P^2}{n_0}-1,
\end{equation}
where $\lambda_e=c/\omega_{pe}$ is the electron skin depth. Finally, solving for $P^2$ and inserting the
result into Eq. (\ref{A}), we obtain
\begin{equation}
  \lambda_e^2\frac{d^2 A}{d \xi^2}+\frac{\omega_0^2}{\omega_p^2} A=
  \left(\lambda_e^2 \frac{d^2\gamma_A}{d \xi^2}+1\right)\frac{A}{\gamma_A},
  \label{A2}
\end{equation}
where we have used $v_0=0$ and, therefore, $k_0=0$. Equation (\ref{A2}) is identical to the model of
Marburger and Tooper \cite{Marburger75} for the nonlinear optical standing wave in a classical
cold fluid electron plasma. The relativistic mass increase is reflected by the ratio ${A}/{\gamma_A}$
in the right-hand side of Eq. (\ref{A2}). The nonlinear electron density fluctuations, which are reflected
in the term proportional to ${d^2\gamma_A}/{d \xi^2}$ in the right-hand side of Eq. (\ref{A2}) can
often be neglected in the weakly relativistic case \cite{Shukla86}.

We note that our previous model \cite{Shukla07} can be recovered in the weakly relativistic limit, in the
following manner. Assuming that, to first order, we have a balance between the ponderomotive and
electrostatic pressure so that $1+ e\phi/m_e c^2\approx \gamma_A$ and that $\gamma_A \approx 1$, and
$v_0^2\ll c^2$. Accordingly, we have
\begin{equation}
  \hbar^2 c^2 \frac{d^2 P}{d\xi^2}+2 m_e^2c^4\bigg(1+\frac{e\phi}{m_e c^2}-\gamma_A\bigg)P=0,
\end{equation}
Poisson's equation
\begin{equation}
  \frac{d^2\phi}{d\xi^2}=\frac{e n_0}{\varepsilon_0}\bigg(\frac{\gamma_A P}{n_0}-1\bigg),
\end{equation}
and the CPEM wave equation
\begin{equation}
  \frac{d^2 A}{d\xi^2}+\lambda A=\frac{\omega_{pe}^2}{c^2}\bigg(\frac{P^2}{n_0} -1 \bigg)A.
\end{equation}

We now make a simple change of variables $\gamma_A P=\widetilde{P}$. Then, we have,
by neglecting terms containing $d\gamma_A/d\xi$, the model \cite{Shukla07}
\begin{equation}
  \hbar^2 c^2 \frac{d^2 \widetilde{P}}{d\xi^2}+2 m_e^2c^4\bigg(1+\frac{e\phi}{m_e c^2}
-\gamma_A\bigg)\widetilde{P}=0,
\end{equation}
\begin{equation}
  \frac{d^2\phi}{d\xi^2}=\frac{e n_0}{\varepsilon_0}\bigg(\frac{\widetilde{P}}{n_0}-1\bigg),
\end{equation}
and
\begin{equation}
  \frac{d^2 A}{d\xi^2}+\lambda A=\frac{\omega_{pe}^2}{c^2}\bigg(\frac{\widetilde{P}^2}{\gamma_A n_0} -1 \bigg)A,
\end{equation}
where the relativistic mass increase appears explicitly in the CPEM wave equation.

\section{Nonlinear dynamics of interacting electromagnetic waves in quantum plasmas}

\begin{figure}[htb]
\includegraphics[width=8.5cm]{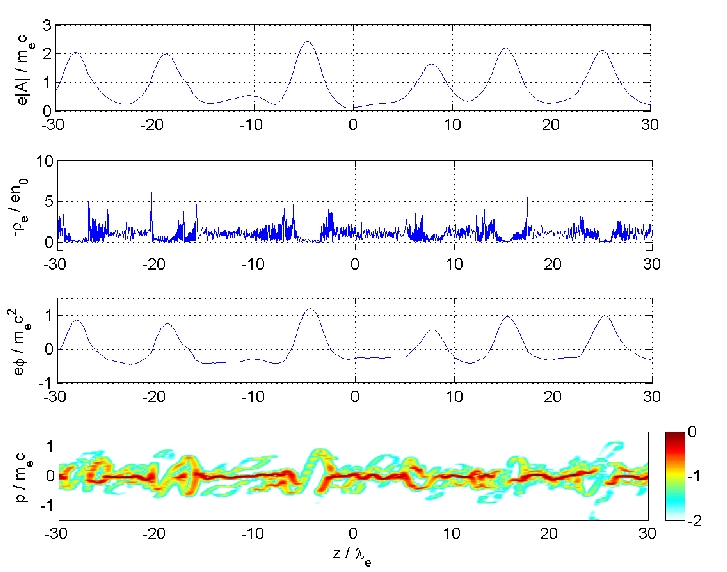}
\caption{The nonlinear stage of a modulational instability, showing localized
electromagnetic wave-packets associated with depletions in the electron charge density,
and positive electrostatic potentials. The electron pseudo-distribution function shows
that the electrons have been accelerated and form BGK-like modes that travel away from
the collapsed wavepackets. Parameters are $H=0.007$, and initially a dipole field $a_0=1$ with $k_0=0$.}
\label{fig:modulational}
\end{figure}

In order to study the dynamics of the nonlinear interaction between
intense CPEM waves and a quantum plasma, we have carried
out numerical simulations of the Klein-Gordon-Maxwell system of equations.
We have here restricted our study to one-space dimension, along
the $z$ direction in space, and written our governing equations in the form
\begin{align}
&\left(i\hbar\frac{\partial}{\partial t}+e\phi\right)\psi=W,
\\
\begin{split}
  &\left(i\hbar\frac{\partial}{\partial t}+e\phi\right)W+\hbar^2 c^2 \frac{\partial^2\psi}{\partial z^2}
  -\gamma_A^2 m_e^2 c^4\psi=0,
\end{split}
\label{KG_1D}
\end{align}
\begin{equation}
\frac{1}{c^2}\frac{\partial^2{\bf A}}{\partial t^2}-\frac{\partial^2{\bf A}}{\partial z^2}
=-\frac{\mu_0 e^2}{m_e} |\psi|^2 {\bf A} ,
\label{vector_1D}
\end{equation}
and
\begin{equation}
  \frac{\partial^2\phi}{\partial z^2}=\frac{e}{2 m_e c^2\varepsilon_0}(\psi^\ast W+\psi W^\ast)
-\frac{e n_0}{\varepsilon_0}.
  \label{Poisson_1D}
\end{equation}
We used a periodic simulation box in space, of length $L_x=63\,\lambda_e$ and used of the order $10^4$ grid points
to resolve the solution in space. It is important to resolve the relatively long electron skin depth
scale as well as the shorter length scale associated with accelerated electrons with
the momentum $p=\hbar k$ and the associated wavelength $\lambda=2\pi/k=2\pi\hbar/p$.
Since we need at least two grid-points per wavelength to represent the solution, the
required grid size is $\Delta x < \pi \hbar/p$, which can be written
$\Delta x/\lambda_e<\pi H m_e c /p$.
For example, to represent the wave function of relativistic electrons with
the momentum $p=m_e c$, we need a spatial grid with
$\Delta x/\lambda_e<\pi H\approx 0.022$ to represent the wave function for $H=0.007$.
The solution was advanced in time with the standard
4th-order Runge-Kutta scheme, using a timestep of order $\Delta t=10^{-4}\,\omega_{pe}$.

\begin{figure}[htb]
\includegraphics[width=8.5cm]{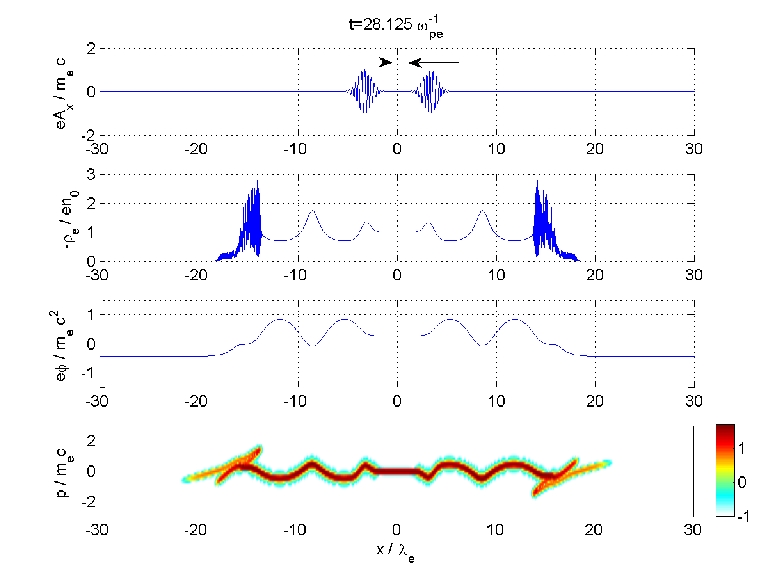}
\caption{Attosecond laser pulse propagation into an underdense quantum plasma, at time $t=28.125\,\omega_{pe}^{-1}$.
Top to bottom panels show a) the electromagnetic vector potential of the laser pulse (the arrows show the
propagation directions of the pulses), b) the electron charge density, c) the electrostatic potential,
and d) the distribution of electrons in phase space in a 10-logarithmic color scale.  Parameters are $H=0.007$,
amplitude $a_0=1$ and wavenumber $k_0=20\,\lambda_e^{-1}$. The laser pulses excite large amplitude oscillatory
potential wakes behind them, as they penetrate the plasma slab.}
\label{fig:colliding1}
\end{figure}

\begin{figure}[htb]
\includegraphics[width=8.5cm]{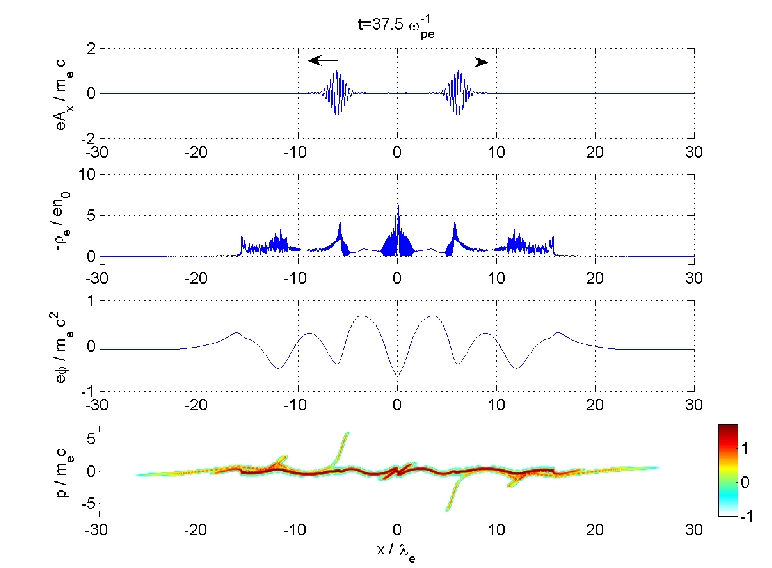}
\caption{The same as in Fig. \ref{fig:colliding1} at time $t=37.5\,\omega_{pe}^{-1}$. Groups of electrons
are accelerated to ultra-relativistic speeds by the large amplitude electrostatic wake field.}
\label{fig:colliding2}
\end{figure}

We first study the growth and nonlinear saturation of the modulational instability,
which is relevant for dense matters where the plasma is overdense or close to overdense.
As initial conditions, we used a CPEM pump wave of the form ${\bf A}=A_0 [\widehat{\bf x}\cos(k_0 z)
-\widehat{\bf y}\sin(k_0 z)]$ with $k_0=0$ and $A_0/m_e c=1$. A small amplitude noise (random numbers)
was added to $A$ in order to seed any instability in the system.
As initial conditions for the wave function,
we used $\psi=\sqrt{n_0/\gamma_A}$ and $W=m_e c^2\sqrt{n_0\gamma_A}$, corresponding
to a pure electronic state at equilibrium. The initially homogeneous
electron density was set to $n_e=n_0=10^{30}\,\mathrm{m}^{-3}$,
corresponding to $H=0.007$ [cf. Eq. (\ref{H})].
In this situation, the electromagnetic wave is unstable due to the
modulational instability, which has instability for small wavenumbers, as
shown in Fig. 3. In the nonlinear stage, the EM waves self-focus into
localized wavepackets similar to the ones in Fig. 4.
Figure \ref{fig:modulational} depicts the late stage of the modulational instability.
The collapse of the CPEM wave packet leads to relativistically strong
ponderomotive potentials that accelerate the electrons to relativistic
speeds. The relativistic electrons are associated with small-scale
spatial oscillations in the wave function, where the wavelength is
comparable to or even smaller than the Compton wavelength.
We see in Fig. \ref{fig:modulational} that the CPEM wave envelope has been focused into
localized wavepackets, associated with depletions in the electron density
and positive localized electrostatic potentials. In order to study the distribution
of electrons both in space and momentum space, we have performed a Fourier transform
of the wavefunction $\psi$ using a moving window technique (using a Hann window) in space.
The width of the window has been tuned so that it provides a good resolution both in
space and in momentum space. The resulting spatial spectrogram gives a representation
of the distribution of the electrons both in space and in momentum space; see
Fig. \ref{fig:modulational}(d), where the color indicates the density of electrons in phase space.
In Fig. 6(d) the horizontal axis shows the spatial dependence and the vertical axis shows the
momentum dependence via the relation $p=\hbar k$ between the momentum $p$ and the wavenumber $k$.
In this figure, it is clear that in the collapse stage of the solitary waves, bunches of electrons
are accelerated to relativistic speeds and form self-trapped, Bernstein-Greene-Kruskal (BGK)-like modes
that propagate away from the collapsed electromagnetic wavepackets.

Next, we investigate a scenario of the short EM pulse propagation and the wake-field generation in a
quantum plasma. This concept is traditionally used for the electron acceleration in classical plasmas
\cite{Tajima79,Bingham04}. The numerical results are displayed in Figs. \ref{fig:colliding1} and \ref{fig:colliding2}.
Here, two atto-second pulses are injected from each side of a plasma slab and are allowed to collide at the
center of the slab. As initial conditions, we used a CPEM pump wave of the form
${\bf A}=A_0(z) [\widehat{\bf x}\cos(k_0 z)-\widehat{\bf y}\sin(k_0 z)]$ with $k_0=20\,\lambda_e^{-1}$
and envelopes of the form $A_0(z)/m_e c=\exp(-(z/\lambda_e\pm 30)^2)$ propagating into the plasma slab.
The plasma slab is initially centered between $z=\pm 15\lambda_e$ with equal electrons with the number
densities $n_0$, where the electron wave function was set to $\psi=\sqrt{n_0}$ and $W=m_e c^2\sqrt{n_0}$.
After a time $t=28.125\,\omega_{pe}^{-1}$, we see in Fig. \ref{fig:colliding1}(b) and (c) that
the large amplitude CPEM pulses excite plasma wake oscillations associated with large-amplitude positive
potentials, and with an approximate wavelength of $2\pi c/\omega_{pe}$, corresponding to a leading
wavenumber of $\omega_{pe}/c$. The positive potentials of the plasma wake oscillations are starting
to capture populations of the electrons at edges of the plasma slab, at $x\approx \pm 15 \lambda_e$.
A high-frequency diffraction pattern is formed in the electron density, as faster electrons overtake slower electrons.
Later, at $t=37.5\,\omega_{pe}^{-1}$, the two laser pulses have collided and passed through each other.
The trapped electrons have been further accelerated to ultra-relativistic speeds, as seen in
Fig. \ref{fig:colliding2}(d) at $x\approx \pm 5 \lambda_e$.,
where the fastest electrons have reached a momentum of $\approx 5\,\mathrm{m_e c}$.

\section{Summary and conclusions}

In this paper, we have developed a relativistic model for the interaction between
intense electromagnetic waves and a quantum plasma. Our nonlinear model
is based on the coupled Klein-Gordon and Maxwell equations for the relativistic
electron momentum and the electromagnetic fields. In our fully relativistic model,
the electron current and charge densities are calculated self-consistently
from the KGE, and they enter as sources for the nonlinear EM and electrostatic waves
in the Maxwell equation. The KG-Maxwell system of equations has been used to derive the
linear dispersion relation for the electrostatic and electromagnetic waves, as well as for
investigating stimulated  Raman scattering and modulational instabilities in the presence
of relativistically intense CPEM waves. In the linear regime, the general dispersion relation
for the electrostatic waves exhibits the quantum effect associated with the overlapping wave
function. At long-wave-lengths, we have the dispersive Langmuir waves with
frequencies close to the electron plasma frequency, while at shorter-wavelengths,
we have the oscillation frequency of free electrons. At wavelengths comparable to
or larger than the Compton wavelength, the electron motion is fully relativistic.
In the nonlinear regime, we have demonstrated the existence of fully relativistic stimulated
Raman scattering and modulational instabilities. While the Raman amplification is
of much interest for generating a coherent electromagnetic radiation, the modulational
instability gives rise to the localization and collapse of the CPEM waves into localized
solitary EM wavepackets. Indeed, numerical simulations of the coupled KG-Maxwell equations
reveal the collapse and acceleration of the electrons in the nonlinear stage of the modulational
instability, as well as the possibility of wake-field acceleration of the electrons to relativistic
speeds by short laser pulses at nanometer scales.  In conclusion, we stress that the present
investigation of nonlinear effects dealing with intense EM wave interactions with quantum plasmas
is relevant for the compression of X-ray free-electron laser pulses to attosecond
duration \cite{Paul01,Trines10}, as well as to the understanding of localized intense
X-ray and $\gamma$-ray bursts that emanate from compact astrophysical objects \cite{Coe,Hurley}.

\acknowledgments
This work was financially supported by the Swedish Research Council (VR).


\begin{thebibliography}{99}
  \bibitem{Hand09} E. Hand, Nature (London) {\bf 461}, 708 (2009).
  \bibitem{Glenzer} S. H. Glenzer {\it et al.}, Phys. Rev. Lett. {\bf 98}, 065002 (2007);
P. Neumayer {\it et al.}, {\it ibid.} {\bf 105}, 075003 (2010); S. H. Glenzer and R. Redmer,
Rev. Mod. Phys. {\bf 81}, 1625 (2009).
  \bibitem{Andreev} A. V. Andreev, JETP Lett. {\bf 72}, 238 (2000).
  \bibitem{Bulanov}G. Mourou {\it et al.}, Rev. Mod. Phys. {\bf 78}, 309 (2006).
  \bibitem{MarklundShukla} M. Marklund and P. K. Shukla, Rev. Mod. Phys. {\bf 78}, 591 (2006).
  \bibitem{Drake74} J. F. Drake, P. K. Kaw, Y. C. Lee, G. Schmidt, C. S. Liu, and M. N. Rosenbluth,
   Phys. Fluids {\bf 17}, 778 (1974).
\bibitem{Sharma83} R. P. Sharma and P. K. Shukla, Phys. Fluids {\bf 26}, 87 (1983).
\bibitem{Murtaza84} G. Murtaza and P.  K. Shukla, J. Plasma Phys, {\bf 31}, 423 (1984).
\bibitem{Shukla} P.K. Shukla {\it et al.}, Phys. Rep. {\bf 138}, 1 (1986).
  \bibitem{McKinstrie89} C. J. McKinstrie and R. Bingham, Phys. Fluids B {\bf 1}, 230 (1989).
  \bibitem{Tsintsadze91} L. N. Tsintsadze, Sov. J. Plasma Phys. {\bf 17}, 872 (1991).
  \bibitem{McKinstrie92} C. J. McKinstrie and R. Bingham, Phys. Fluids B {\bf 4}, 2626 (1992).
  \bibitem{Sakharov94} A. S. Sakharov and V. I. Kirsanov, Phys. Rev. E {\bf 49}, 3274 (1994).
  \bibitem{Guerin95} S. Gu\'erin, G. Laval, P. Mora, J. C. Adam, A. H\'eron, and A. Bendib,
Phys. Plasmas {\bf 2}, 2807 (1995).
  \bibitem{Adam00} J. C. Adam, A. H\'eron, G. Laval, and P. Mora, Phys. Rev. Lett. {\bf 84}, 3598 (2000).
  \bibitem{Quesnel97} B. Quesnel, P. Mora, J. C. Adam, A. H\'eron, and G. Laval,
Phys. Plasmas {\bf 4}, 3358 (1997).
  \bibitem{Stenflo76} L. Stenflo, Phys. Scr. {\bf 14}, 320 (1976).
  \bibitem{Stenflo80} L. Stenflo, Phys. Scr. {\bf 21}, 831 (1980).
  \bibitem{Stenflo81} L. Stenflo and H. Wilhelmsson, Phys. Rev. A {\bf 24}, 1115 (1981).
  \bibitem{Malkin07} V. M. Malkin, N. J. Fisch, and J. S. Wurtele,  Phys. Rev. E {\bf 75}, 026404 (2007).
  \bibitem{Serbeto08} A. Serbeto, J. T. Mendon\c{c}a, K. H. Tui {\it et al.}, Phys.
Plasmas {\bf 15}, 013110 (2008).
  \bibitem{Serbeto09} A. Serbeto, L. F. Monteiro, K. H. Tsui, and J. T. Mendon\c{c}a,
Plasma Phys. Control. Fusion {\bf 51} 124024 (2009).
  \bibitem{Piovella08} N. Piovella, M. M. Cola, L. Volpe, A. Schiavi, R. Bonifacio,
  Phys. Rev. Lett. {\bf 100}, 044801 (2008).
  \bibitem{Chabrier} G. Chabrier {\it et al.}, J. Phys.: Condens. Matter {\bf 14}, 9133 (2002);
  J. Phys. A: Math. Gen. {\bf 39}, 4411 (2006).
  \bibitem{Coe} M. J. Coe {\it et al.}, Nature (London) {\bf 272}, 37 (1978);
    D. K. Galloway and J. L. Sokoloski, Astrophys. J. {\bf 613}, L61 (2004).
  \bibitem{Hurley} K. Hurley {\it et al.}, Nature (London) {\bf 434}, 1098 (2005);
    A. K. Harding and D. Lai, Rep. Prog. Phys. {\bf 69} 2631 (2006).
  \bibitem{Manfredi01} G. Manfredi and F. Haas, Phys. Rev. B {\bf 64}, 075316 (2001);
P. K. Shukla and B. Eliasson, Phys. Rev. Lett. {\bf 96}, 245001 (2006); Phys. Usp. {\bf 53}, 51 (2010).
  \bibitem{Fuda82} M. G. Fuda and E. Furlani, Am. J. Phys. {\bf 50}, 545 (1982).
  \bibitem{Gerritsma10} R. Gerritsma {\it et al.}, Nature (London) {\bf 463}, 68 (2010).
  \bibitem{Akhiezer56} A. I. Akhiezer and R. V. Polovin, Sov. Phys. JETP {\bf 3}, 696 (1956)
[Zh. Eksp. Teor. Fiz. {\bf 30}, 915 (1956)]; P. Kaw and J. Dawson, Phys. Fluids {\bf 13}, 472 (1970);
C. Max and F. Perkins, Phys. Rev. Lett. {\bf 27}, 1342 (1971).
  \bibitem{Marburger75} J. H. Marburger and R. F. Tooper, Phys. Rev. Lett. {\bf 35}, 1001 (1975).
  \bibitem{Gersten75} J. I. Gersten and N. Tzoar, Phys. Rev. Lett. {\bf 35}, 934 (1975).
  \bibitem{Shukla07} P. K. Shukla and B. Eliasson, Phys. Rev. Lett. {\bf 99}, 096401 (2007).
  \bibitem{Kaw92} P. K. Kaw, A. Sen, and T. Katsouleas, Phys. Rev. Lett. {\bf 68}, 3172 (1992).
  \bibitem{Saxena06} V. Saxena, A. Das, A. Sen, and P. Kaw, Phys. Plasmas {\bf 13}, 032309 (2006).
  \bibitem{Shukla86} P. K. Shukla, N. N. Rao, M. Y. Yu, and N. L. Tsintsadze, Phys. Rep. {\bf 138}, 1 (1986).
  \bibitem{Tajima79} T. Tajima and J. M. Dawson, Phys. Rev. Lett. {\bf 43}, 267 (1979).
  \bibitem{Bingham04} R. Bingham {\it et al.}, Plasma Phys. Control.  Fusion {\bf 46}, R1 (2004).
\bibitem{Paul01} P. M. Paul {\it et al.}, Science {\bf 292}, 1689 (2001).
\bibitem{Trines10} R. M. G. M. Trines, F. Fiuza, R. Bingham {\it et al.}, Nature Phys. {\bf 6}, 1793
doi:10.1038/nphys1793 (2010).
\end{thebibliography}
\end{document}